\documentclass{sigchi-ext}
\usepackage{calc}
\usepackage[T1]{fontenc}
\usepackage{textcomp}
\usepackage[scaled=.92]{helvet} 
\usepackage{graphicx} 
\usepackage{balance}  
\usepackage{booktabs} 
\usepackage{}  
\usepackage{ragged2e} 

\copyrightinfo{Permission to make digital or hard copies of part or all of this work for personal or classroom use is granted without fee provided that copies are not made or distributed for profit or commercial advantage and that copies bear this notice and the full citation on the first page. Copyrights for third-party components of this work must be honored. For all other uses, contact the Owner/Author. \\
Copyright is held by the owner/author(s).\\
\emph{CHI'16 Extended Abstracts}, May 7--12, 2016, San Jose, CA, USA.\\
ACM 978-1-4503-4082-3/16/05.\\
http://dx.doi.org/10.1145/2851581.2892466}

\copyrightinfo{\copyright~Scott A.~Hale 2016. This is the author's version of the work. It is posted here for your personal use. Not for redistribution. The definitive version was published in CHI EA 2016, http://dx.doi.org/10.1145/2851581.2892466.}

\numberofauthors{1}
\author{%
  \alignauthor{%
    \textbf{Scott A.~Hale}\\
    \affaddr{Oxford Internet Institute} \\
    \affaddr{University of Oxford}\\
    \affaddr{1 St Giles, Oxford, OX1 3JS, UK} \\
    \affaddr{scott@hale.us}%
    } }

\def\plaintitle{User Reviews and Language: How Language Influences Ratings}
\def\plainauthor{Scott A. Hale}
\def\plainkeywords{user-generated content; product reviews; e-commerce; multilingualism; internationalization and localization}

\title{\plaintitle}

\hypersetup{%
  pdftitle={\plaintitle}, pdfauthor={\plainauthor},
  pdfkeywords={\plainkeywords}, }

\teaser{
	\vspace{-8ex}
	\hspace{-.8\marginparwidth}
	\includegraphics[width=\columnwidth+0.8\marginparwidth]{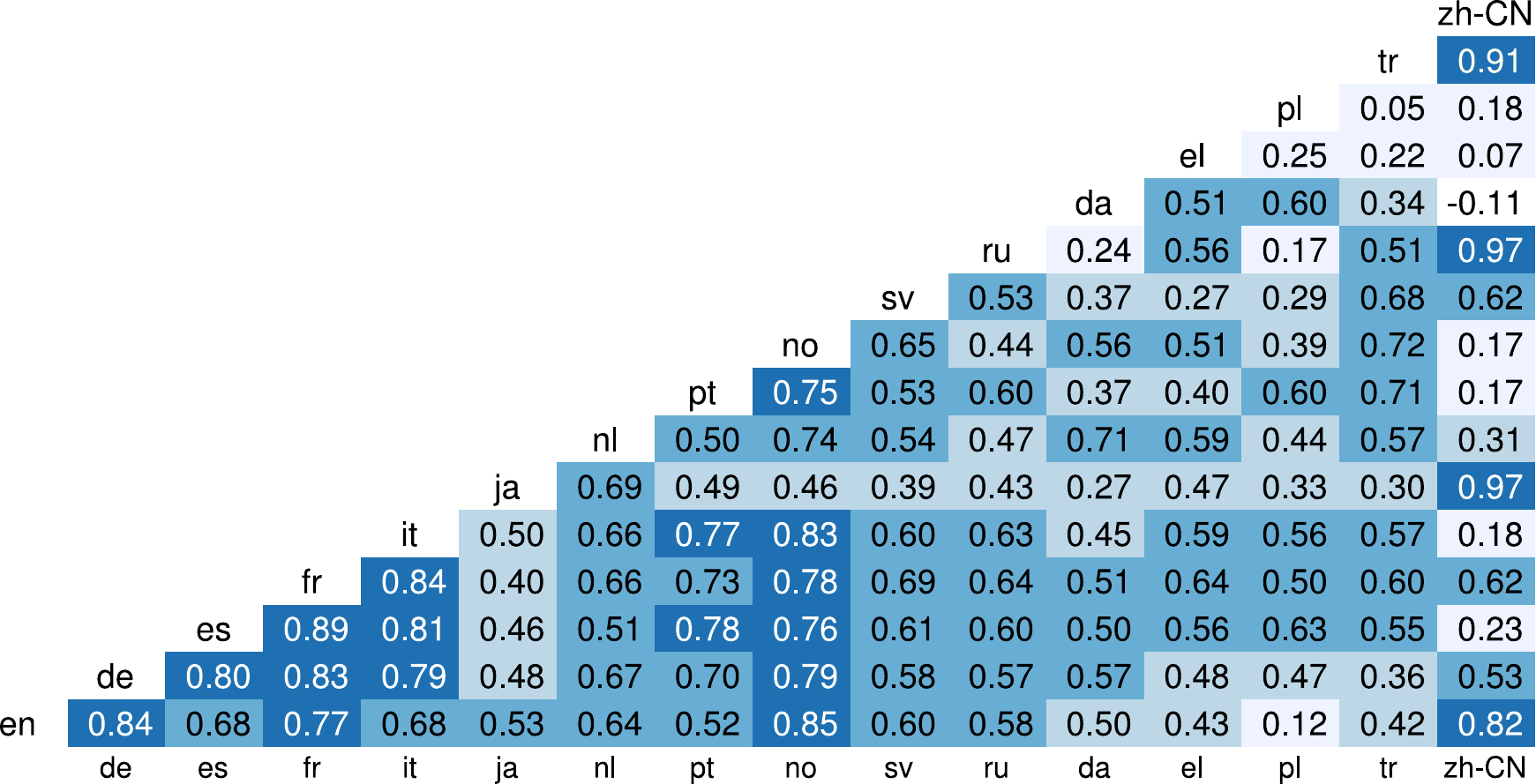}
	\caption{The similarity with which users writing in different languages rated the same London attractions on TripAdvisor. Each cell gives the mean correlation of the star ratings between a pair of languages.}
	\label{fig:lang_sim}
	\vfill 
}

\begin{document}

\maketitle

\RaggedRight{}

\begin{abstract}
  The number of user reviews of tourist attractions, restaurants, mobile apps, etc. is increasing for all languages; yet, research is lacking on how reviews in multiple languages should be aggregated and displayed. Speakers of different languages may have consistently different experiences, e.g., different information available in different languages at tourist attractions or different user experiences with software due to internationalization\slash{}localization choices. This paper assesses the similarity in the ratings given by speakers of different languages to London tourist attractions on TripAdvisor. The correlations between different languages are generally high, but some language pairs are more correlated than others. The results question the common practice of computing average ratings from reviews in many languages.
\end{abstract}



\section{Introduction}
The amount of content online in different languages is greatly increasing, and the early days of English-language dominance on the Web have given way to language pluralism online. For many large user-generated content platforms, less than half the content is in English~\cite{hecht2010,hong2011} and many users do not speak English as a native language~\cite{hale2014twitter,hale2014wiki}. As Internet-penetration rates are already high in most English-speaking countries, future user growth (and the content contributed by these users) will be predominantly in non-English languages \cite{graham-inet-penetration}.

The language dynamics of online reviews have received little scholarly attention, and industry practices vary greatly. In general, many websites aggregate reviews from multiple languages together to compute an average rating as is the case with TripAdvisor, the travel review website analyzed in this paper. Many websites differ, however, on how reviews in other languages are displayed (if at all) to users. TripAdvisor generally shows reviews in reverse chronological order (most recent reviews first), but demotes foreign-language reviews so that they appear after all reviews in the language selected by the user. By contrast, Google Play, a mobile app store, hides reviews in other languages entirely making them completely inaccessible (although reviews from all languages appear to be used when calculating the average rating of an app).\footnote{For more details see, ``Design for multilinguals: Seemingly simple yet often missed,'' \url{http://www.scotthale.net/blog/?p=412}}
Beyond reviews, Twitter, Facebook, and Google Plus all provide the option to see machine translations of foreign-language posts, and Facebook has experimented with showing machine translations in place of foreign-language posts.

In general, a larger number of reviews is thought to be more helpful to potential consumers making purchasing decisions \cite{hu2008,park2007}.
There remains, however, a fundamental question of whether reviews in different languages are analytically similar to each other. If speakers of different languages focus on different aspects, evaluate products differently, and/or have consistently different experiences (e.g., different internationalization\slash{}localization choices for software or different information, etc. available for in-person activities) the reviews from one language may have less relevance to individuals primarily speaking a different language. If so, the practice of creating an average rating from reviews in multiple languages could be unhelpful or even misleading.

\section{Data and methods}
One large segment of user reviews is travel reviews. Tourist attractions in popular, international cities are reviewed by users from many countries, speaking many languages. TripAdvisor is one of the largest platforms for travel reviews, reporting 315 million unique visitors per month.\footnote{\url{http://www.tripadvisor.co.uk/PressCenter-c6-About_Us.html}} All of the reviews on tripadvisor.co.uk about tourist attractions in London, England, were crawled and extracted using a custom-built webcrawler in Python3.\footnote{The code is freely available under an open-source license at \url{http://www.scotthale.net/pubs/?chi2016}.} London is a suitable choice as it is a large, international city and a top tourist destination for people from many countries. At the time of crawling in July 2015, TripAdvisor had 516,641 reviews pertaining to 3,040 different tourist attractions in London. The dataset only includes tourist attractions (as defined by TripAdvisor) and does not include reviews of hotels and restaurants.

TripAdvisor provides a link to machine translate non-English reviews, and the source-language parameter included in that machine translation link was taken as the language of the review. Reviews without a machine translation link were assumed to be in English. A human examination of 100 randomly chosen reviews did not find any errors in language labels. All reviews were also examined with the Compact Language Detection kit used within Google Chrome, and CLD detected the same language as that extracted from the translation link for 99.5\% of the reviews. Ad hoc examination suggested the disagreements between CLD and TripAdvisor were often due to the mixing of multiple languages within a single review.

The name and postcode of each attraction were recorded, and then the following elements for each review of the attraction were extracted:
\begin{itemize}
	\item A numeric id of the user authoring the review stored in the HTML of the page
	\item The ``star'' rating the user gave the attraction. This is a whole number between 1 (the lowest rating) and 5 (the highest rating)
	\item The date on which the user authored the review
	\item The location of the author (free-text, optional)
\end{itemize}
\section{Results}
\begin{figure}
\includegraphics[width=\columnwidth]{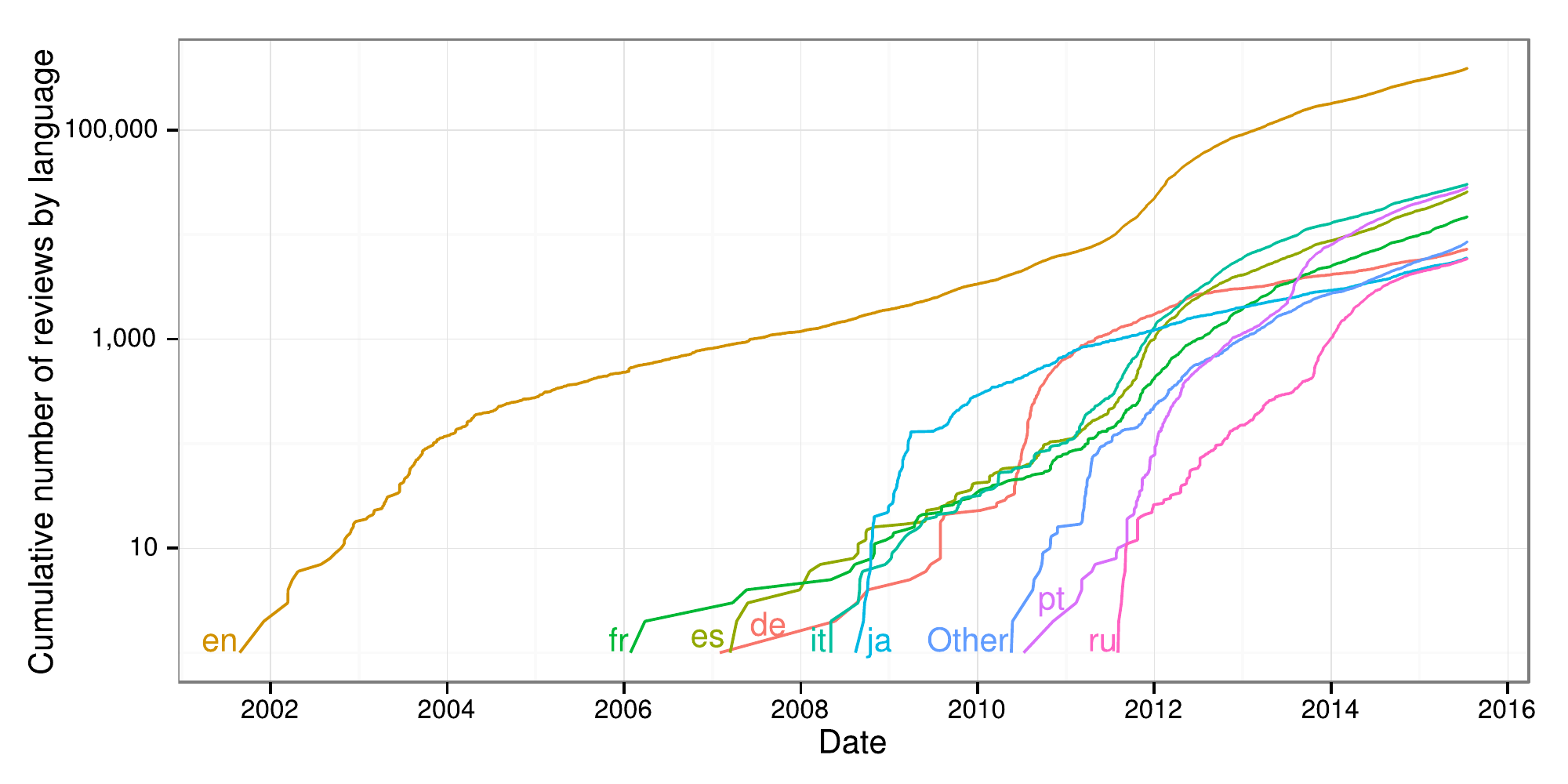}
\caption{The number of user reviews on TripAdvisor about London attractions from 2001 to 2015 for the top 8 languages}
\label{fig:lang_growth}
\end{figure}

The earliest reviews on TripAdvisor date from 2001 and are all in English. However, from 2006 onwards non-English reviews grew quickly as shown in Figure~\ref{fig:lang_growth}. By July 2015 when the site was crawled, 25\% of all reviews of London attractions were not in English. Just over half of all attractions had at least one non-English review, and 175 attractions (6\%) had more non-English than English-language reviews.

\begin{figure}
\hspace{-.5\marginparwidth}%
\includegraphics[width=\columnwidth+0.5\marginparwidth]{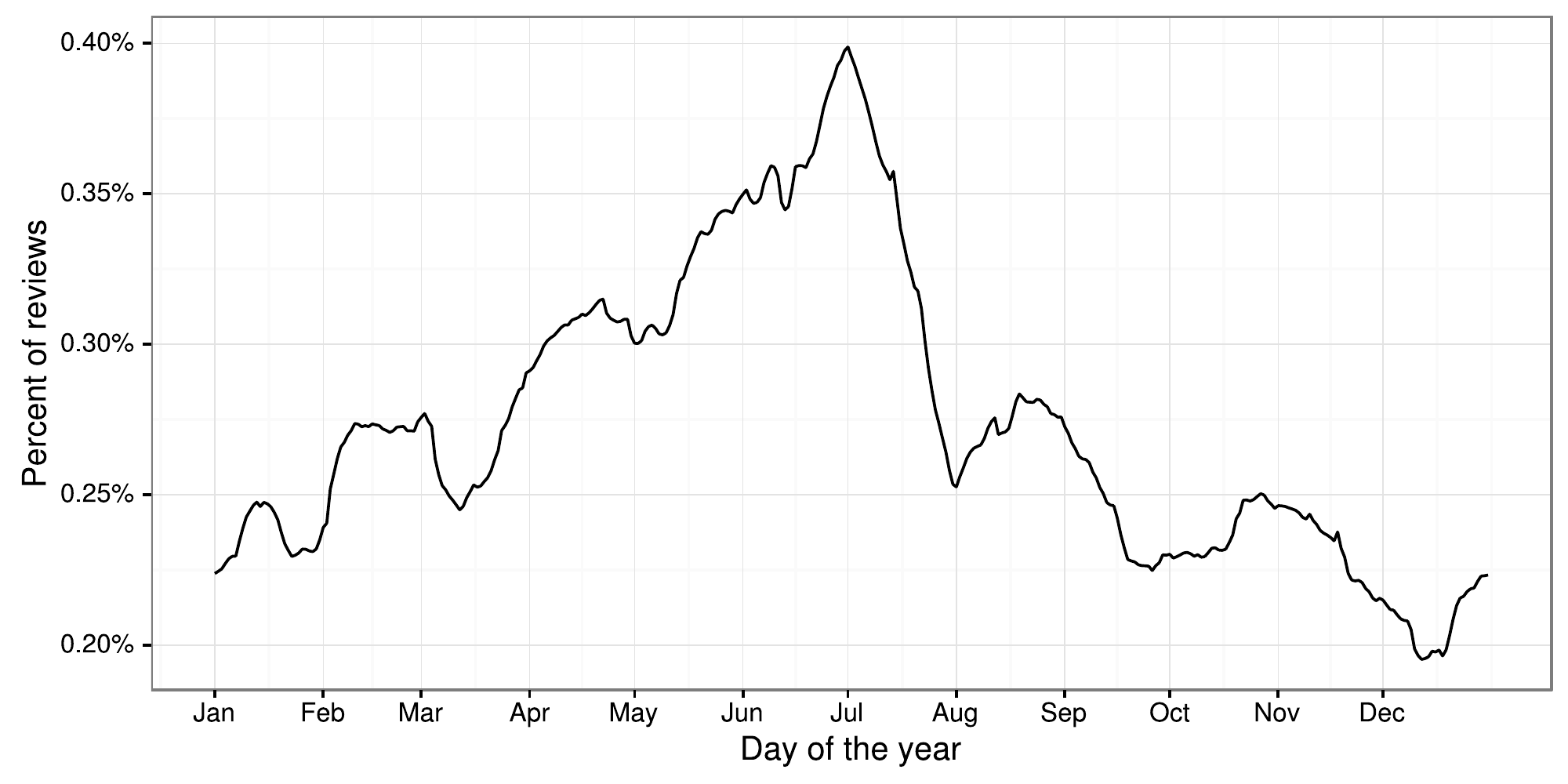}
\caption{The number of user reviews on TripAdvisor about London attractions by day of the year for all languages. The plot combines data from 2001 to 2015, and the data is smoothed with a 30-day rolling window.}
\label{fig:global_doy}
\end{figure}

\begin{figure}
\hspace{-.5\marginparwidth}%
\includegraphics[width=\columnwidth+0.5\marginparwidth]{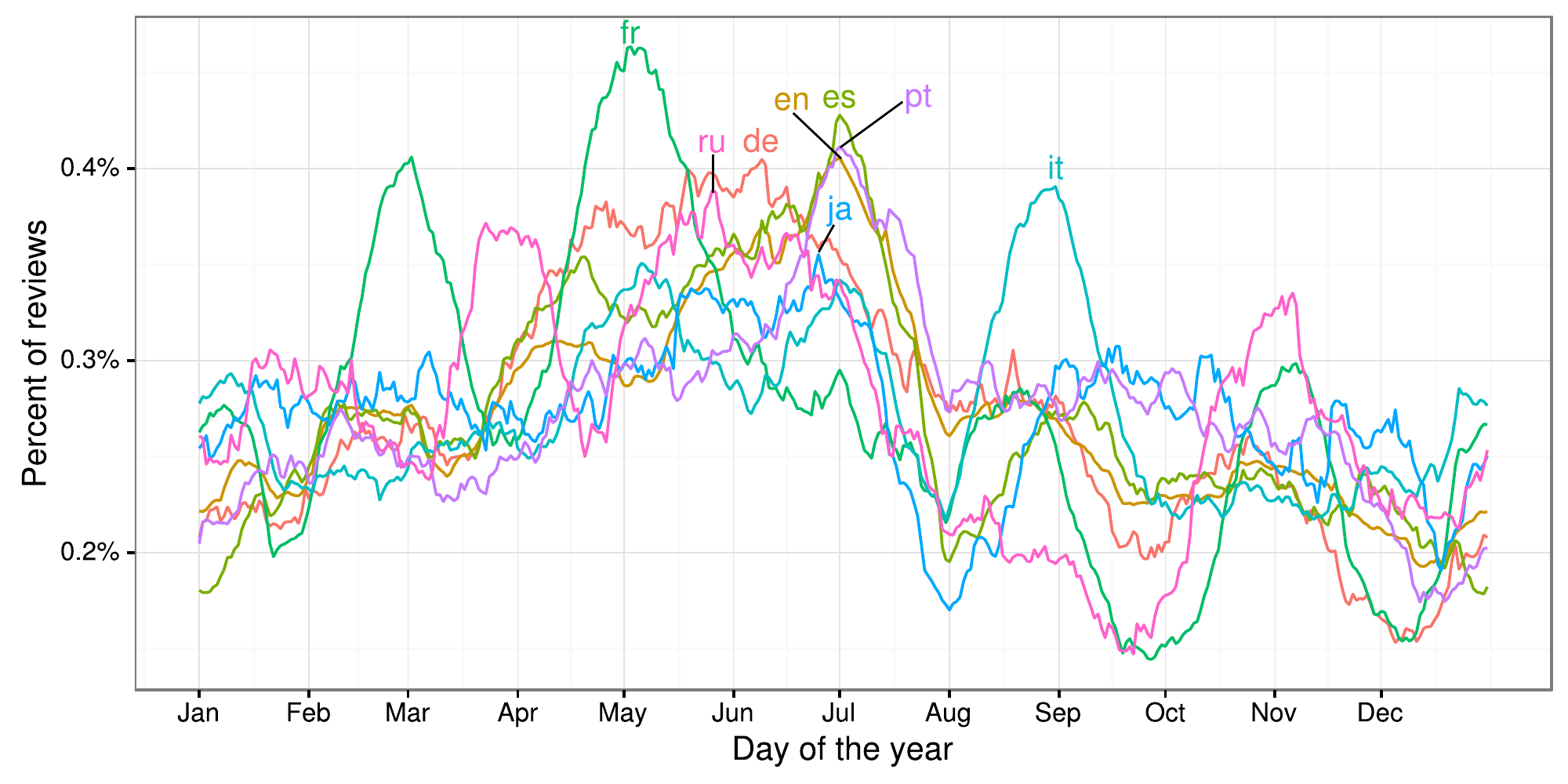}
\caption{The number of user reviews on TripAdvisor about London attractions by day of the year for the top 8 languages. The plot combines data from 2001 to 2015, and the data is smoothed with a 30-day rolling window.} 
\label{fig:doy}
\end{figure}

Reviews of London attractions in all languages tended to be written in the summer months. Using a 30-day rolling window, the window with the most reviews was centered on July 1 and contained 12\% of all reviews (Figure~\ref{fig:global_doy}).\newpage
The timing of reviews in each language was similar but had slight differences. French reviews were written earlier in the year: 14\% of French reviews were written in the 30-day window centered on May 3. Italian reviews were written later in the year: 12\% of Italian reviews were written in the 30-day window centered on August 31.
Figure~\ref{fig:doy} shows the percentage of reviews written in different language each day of the year smoothed using a 30-day rolling window.

The average star rating (1--5 stars) is sensitive to the number of reviews. With a small number of reviews, a single rating can be over represented. Through manual examination of different thresholds, 10 reviews was chosen as the minimum number of reviews needed to consider an attraction.
There were 471 attractions with at least 10 English and 10 non-English reviews, and among these attractions the correlation in the average rating between English and non-English reviews was strong (0.72). On average, the mean ratings of English reviews were 0.067 of a star lower than the mean ratings of non-English reviews (4.22 vs.~4.29 stars). While this difference is statistically significant at conventional levels ($p<0.03$), the magnitude of the difference is very small.

Applying the same criteria of at least 10 reviews in a language and 10 reviews in all other languages, correlations for each language were computed. Each correlation is the average star ratings of speakers of the language compared to the average star ratings of speakers of all other languages. As can be seen in Table~\ref{tbl:cor}, the correlations vary considerably. Ratings in German, Norwegian, and French are strongly correlated with ratings in other languages. In contrast, ratings in languages such as Portuguese and Japanese are less strongly correlated. Thus, the usefulness of reviews in another language may vary by language.

\begin{table}
\hspace{-.5\marginparwidth}
\begin{tabular}{lrrr@{}l}
	\toprule
	Language & Num.\ attractions & Correlation & Mean difference & \\
	\midrule
	Polish (pl)		&	12	&	0.30	&	0.05	&	\\
	Turkish (tr)		&	8	&	0.42	&	0.04	&	\\
	Greek (el)		&	14		&	0.52	&	0.14	&	\\
	Danish (da)		&	22	&	0.52	&	$-$0.18	&	\\
	Japanese (ja)		&	81	&	0.53	&	$-$0.17	&	***\\
	Portuguese (pt)	&	199	&	0.61	&	0.17	&	***\\
	Swedish (sv)		&	38	&	0.63	&	$-$0.11	&	\\
	Russian (ru)		&	98	&	0.66	&	0.29	&	***\\
	Dutch (nl)		&	47	&	0.70	&	$-$0.09	&	\\
	English (en)		&	471	&	0.72	&	$-$0.07	&	*\\
	Italian (it)			&	216	&	0.73	&	0.02	&	\\
	Chinese (zh-CN)	&	6	&	0.74	&	$-$0.10	&	\\
	Spanish (es)		&	174	&	0.77	&	$-$0.01	&	\\
	French (fr)		&	160	&	0.81	&	$-$0.05	&	\\
	Norwegian (no)	&	18	&	0.86	&	$-$0.13	&	\\
	German (de)		&	110	&	0.88	&	0.02	&	\\
	\bottomrule
\end{tabular}
\caption{Correlations and differences between the mean ratings for attractions given by speakers of a language compared to ratings given by speakers of all other languages. Asterisks indicate the significance of two-sample t-tests on the means of the average ratings (***~$p<0.001$, **~$p<0.01$, *~$p<0.05$).}
\label{tbl:cor}
\end{table}

Looking at pairs of languages, the correlations in the star ratings given by speakers of different languages to attractions ranged from a minimum of $-$0.1 between Chinese and Danish to a maximum of 0.97 between Chinese and Japanese as well as between Chinese and Russian. In general, the correlations were high (Figure~\ref{fig:lang_sim}). Within the distribution of correlations, the first quartile was 0.44, the median 0.56, and the third quartile 0.68.


Most users wrote only one review of a London attraction (162,801 of 254,518 users, or 64\%). Of the users writing multiple reviews, a small number wrote reviews in two different languages (943 of 91,717 users, or 1\%). No users wrote reviews in more than two languages. Although a small percentage of users, taken along with the single reviews that mixed multiple languages together, it is important for interface designers to consider bilingual users (including users who might read reviews in multiple languages but only write reviews in one language). Consistent with findings on Wikipedia \cite{hale2014wiki} and Twitter \cite{hale2014twitter}, users writing reviews in more than one language were more active on TripAdvisor than users writing reviews in only one language. Among users writing at least two reviews, users writing in two different languages authored more reviews than users writing in only one language (5.1 vs.~3.8 reviews per user on average; $p<0.001$).


\section{Discussions}
It is common practice to create one overall rating for a product or item by simply averaging all the available ratings without regard to the location or language of the reviewer. With regards to language and tourist attractions in London, this practice seems to be justified in general, although some language pairs are more strongly correlated than others.

In general, research has suggested that a larger number of reviews is more helpful to a person trying to make a decision about a product \cite{hu2008,park2007}.
Users may derive some utility from the star ratings of reviews in languages they do not read and possibly more from rough machine translations of the review text. At the same time, the experience of reviewers speaking a different language may be a poor indication of the experience the person will actually have with a product/service. As far as London attractions are concerned, the star ratings of reviews in different languages have varying correlations with each other. Ratings in German, Norwegian, and French are more strongly correlated with reviews in other languages than are ratings in Japanese, Portuguese, or Russian. Thus, the usefulness that users have from reviews in other languages likely varies with the languages they speak.

When there are few reviews in a user's language(s), it may be helpful to display reviews in other languages. The correlations between pairs of languages suggest that ratings from some languages will be more indicative of the experience a person speaking a given language will have than ratings from other languages. This may be due to underlying elements of culture that are captured by the language(s) of a person. Research has shown some differences in the use of social media platforms correlate with cultural dimensions measured at the country level \cite{garcia2013}, and similar cultural dimensions may affect the expectations and evaluations of people writing reviews in different languages.

Beyond the similarity of evaluations there is also a user-interface design question about how helpful people perceive reviews written in another language. Experiments are an exciting methodology to directly test how people respond to foreign-language content.

This extended abstract is a first and incomplete step into examining reviews in multiple languages. It is unclear how far the findings related to tourist attractions in London, England, extend to other locations or to other types of reviews.
Further work is needed to analyze other types of reviews such as reviews of mobile apps where the user experiences may vary across languages depending on the international and localization choices made. Even seemingly language-neutral factors such as the well-known 140-character limit on Twitter appear to have different effects on users writing in different languages \cite{liao2015,neubig2013}.
Such research is important, along with other information such as rates of bilingualism, for interface designers to decide which foreign-language reviews to show first (if any) and how ratings from multiple languages should be averaged (if at all).
At the same time, it is important to remember that many Internet users are bilingual \cite{hale2014twitter,hale2014wiki}---perhaps even the majority \cite{birner2005,grosjean2010}---and, designers should allow multilingual users access to content in their multiple languages.

\section{Acknowledgements}
I am grateful for funding support to conduct this research from the John Fell Oxford University Press (OUP) Research Fund as well as the University of Oxford's Economic and Social Research Council (ESRC) Impact Acceleration Account and Higher Education Innovation Fund (HEIF) allocation.
I would also like to thank the UK Arts and Humanities Research Council (AHRC) for funding the original collection of TripAdvisor data as part of a project studying the quality and completeness of the web archive data. 


\bibliographystyle{SIGCHI-Reference-Format}


\end{document}